\documentclass[a4paper,10pt,twocolumn,aps,showpacs,pra]{revtex4-1}
\usepackage{graphicx}
\usepackage{dcolumn}
\usepackage{bm}
\usepackage{slashed}
\usepackage{amssymb}
\usepackage{amsmath}
\usepackage{amsthm}
\usepackage{natbib}
\usepackage{color}
\usepackage{hyperref}

\begin{document}

\title{Entanglement of an alternating bipartition in spin chains: Relation with classical integrable models}

\author{Raul A. Santos}
\email{raul.santosza@biu.ac.il}
\affiliation{Department of Condensed Matter Physics, Weizmann Institute of Science, Rehovot, Israel; 
Department of Physics, Bar-Ilan University, Ramat Gan, 52900, Israel}

\begin{abstract}

We study the entanglement properties of a class of ground states defined by matrix product states, which
are generalizations of the valence bond solid (VBS) state in one dimension. It is shown that the transfer matrix of these 
states can be related to representations of the Temperley-Lieb algebra, allowing an exact computation of Renyi entropy. 
For an alternating bipartition, we find that the Renyi entropy can be mapped to an eight vertex 
model partition function on a rotated lattice. We also show that for the VBS state, the Renyi entropy 
of the alternating partition is described by a critical field theory with central charge $c=1$. 
The generalization to $SU(n)$ VBS and its connection with a dimerization transition in the entanglement
Hamiltonian is discussed.

\end{abstract}


\maketitle

In recent years it has become clear that any attempt for classification of quantum systems should account for their
entanglement properties \cite{Chen2011,Turner2011,Fidkowski2011,Schuch2011}. This approach has been specially fruitful in gapped systems where the 
energy gap induces some robustness against local perturbations. For temperatures well below the gap, the system resides 
in its ground state, making this state specially important in the description of a quantum system.
An efficient way to describe entangled ground states is by Tensor Product States (TPS). They provide a
useful representation of short \cite{Orus2014} and long range entangled states \cite{Verstraete2006PRL,Wahl2013,Dubail2013}.
In one dimension, an specific type of TPS called matrix product states (MPS) provides a way to construct any ground state 
of a gapped system \cite{Verstraete2004,verstraete2006}. 

In this context, a characterization of the ground state entanglement becomes very necessary. 
Among the different measures of entanglement in quantum systems, a particularly useful one is the entanglement 
entropy (EE) (or its generalization as a Renyi entropy). In pure states, like the ground state, EE gives a 
unique measure of the entanglement present between two complementary subsystems.

In this letter we study a class of one dimensional 
spin ground states that generalizes the VBS states. These states, defined by their matrix
product representation, are ground states of parent Hamiltonians which are non integrable. Despite of their
lack of integrability the Renyi entropy of a partition sensitive to the bulk entanglement,
introduced in Ref. \cite{Hsieh2014a}, can be mapped to partition functions of integrable models.

This paper is organized as follows. First we review the construction of the
(VBS) state in one dimension and its MPS, particularly in the context of the AKLT 
\cite{Affleck1987} model. We also review the computation of correlation functions and entanglement measures 
(Renyi and entanglement entropy) in the ground state by means of the transfer matrix technique.
In the second section we propose an MPS that contains the VBS state as a particular limit. For these generalized
states, we show that a particular bipartition, dubbed alternating bipartition (AB), maps naturally to 
the partition function of an eight vertex model. This classical two dimensional model has analytic solution
as its Boltzmann weights satisfy the Yang-Baxter equation \cite{Baxter8v1971}. We find that for the AKLT ground state, 
the corresponding eight vertex model becomes critical. The connection is made by showing that the transfer matrix
of the VBS state is related with a representation of the Temperley-Lieb algebra \cite{TLalgebra}.
This allows further characterization of the AB entanglement as the free energy of a conformal field theory (CFT) with 
central charge $c=1$. 
Considering the generalization of the VBS state for larger symmetry groups $SU(n)$, we find that the
Reyni entropy in this case is described by the partition function of the $n^2$-state Potts model, at the transition
point. This transition is first order for $n>2$, so is not described by a CFT. 
Finally, we discuss the existence of a representative among this generalized MPS whose AB entanglement is given by a 
topological quantity. It corresponds to a sum over loop configurations on the induced two 
dimensional lattice, which depends on the boundary conditions of the spin chain.

\section{AKLT model and Matrix Product Representation}

The AKLT model is considered to be the first example of a matrix product state. It was introduced in \cite{Affleck1987}
in order to study the Haldane conjecture \cite{Haldane1983} for integral spin chains. The model consists of a spin chain 
where at each site of 
the lattice resides a spin 1 particle. The Hamiltonian for this model is the sum of projection operators of nearest 
neighbor spins onto total spin 2. Specifically, for periodic boundary conditions, the Hamiltonian of an $N$-sites chain 
reads

\begin{equation}\label{AKLT}
 H=\frac{1}{2}\sum_{i=1}^{N}\left(\frac{2}{3}+S_i\cdot S_{i+1}+\frac{1}{3}(S_i\cdot S_{i+1})^2\right).
\end{equation}

The ground state of this Hamiltonian is easily found by the VBS construction. 
At each site, spin 1 states are constructed by projecting the tensor product of two spin $\frac{1}{2}$ representations
onto the symmetric subspace. This reproduces the spin 1 Hilbert space at each site. 
The spin $\frac{1}{2}$ representations at adjacent sites are antisymmetrized forming singlets. This prevents
the formation of spin 2 states between nearest neighbors. Hence the total spin between adjacent sites cannot be 
larger than 1 $(|\vec{S}_i+\vec{S}_{i+1}|=0,1)$, 
thus being annihilated by the AKLT Hamiltonian (\ref{AKLT}).
As this Hamiltonian is a sum of projectors, its eigenvalues are semi positive definite, with the ground state having
eigenvalue zero. This shows that the VBS state is the ground state of (\ref{AKLT}).

The matrix product representation of the VBS state has been studied in many contexts (see for example
\cite{Schollwock2011,PerezGarcia2007}). For completeness here we rederive it. It is simply found from the previous discussion as follows.
We define the matrices $A^{[-1]}, A^{[0]}$ and $A^{[1]}$, related with the symmetric subspace of the
tensor product of two spin $\frac{1}{2}$ states by (sum over repeated indices implied, otherwise noted)

\begin{equation}
 \left|\frac{1}{2},a\right\rangle_i\otimes\left|\frac{1}{2},b\right\rangle_i+ \left|\frac{1}{2},b\right\rangle_i\otimes\left|\frac{1}{2},a\right\rangle_i\equiv A_{ab}^{[m]}|1,m\rangle_i,
\end{equation}

\noindent where $|1,m\rangle$ is a state of definite spin 1 and $m=S_z$ . Here each label $a$ and $b$ takes two values $\pm\frac{1}{2}$. Antisymmetrization 
of neighboring sites is achieved by inserting the totally antisymmetric tensor $\epsilon_{ab}$ between them

\begin{equation}
A_{ab}^{[m]}|1,m\rangle_i\epsilon_{bc}A_{cd}^{[m']}|1,m'\rangle_{i+1}.
\end{equation}

\noindent Defining finally $A^{[m]}\epsilon=M^{[m]}$, we find for the VBS state $|\Omega\rangle$
its matrix product representation

\begin{equation}
 |\Omega\rangle={\rm Tr}(M^{[m_1]}M^{[m_2]}\dots M^{[m_N]})|m_1,m_2\dots,m_N\rangle,
\end{equation}

\noindent where we have used the notation $|1,m\rangle\rightarrow|m\rangle$. The
state $|m_1,\dots, m_N\rangle$ is the tensor product over all lattice sites. 
The matrices $M^{[k]}$, written in terms of Pauli matrices $\sigma^{\pm}=\sigma^x\pm i\sigma^y$,
$\sigma^z$ are explicitly (up to an overall factor)

\begin{eqnarray}\nonumber
 M^{[-1]} =-\sqrt{2}\sigma^{-},\quad M^{[1]} =\sqrt{2}\sigma^{+},\quad  M^{[0]} =-\sigma^z.
\end{eqnarray}

The matrices $M^{[k]}$ define a map from an auxiliary space to the physical Hilbert space of spin states.

\begin{figure}[ht!]
\includegraphics[scale=.45]{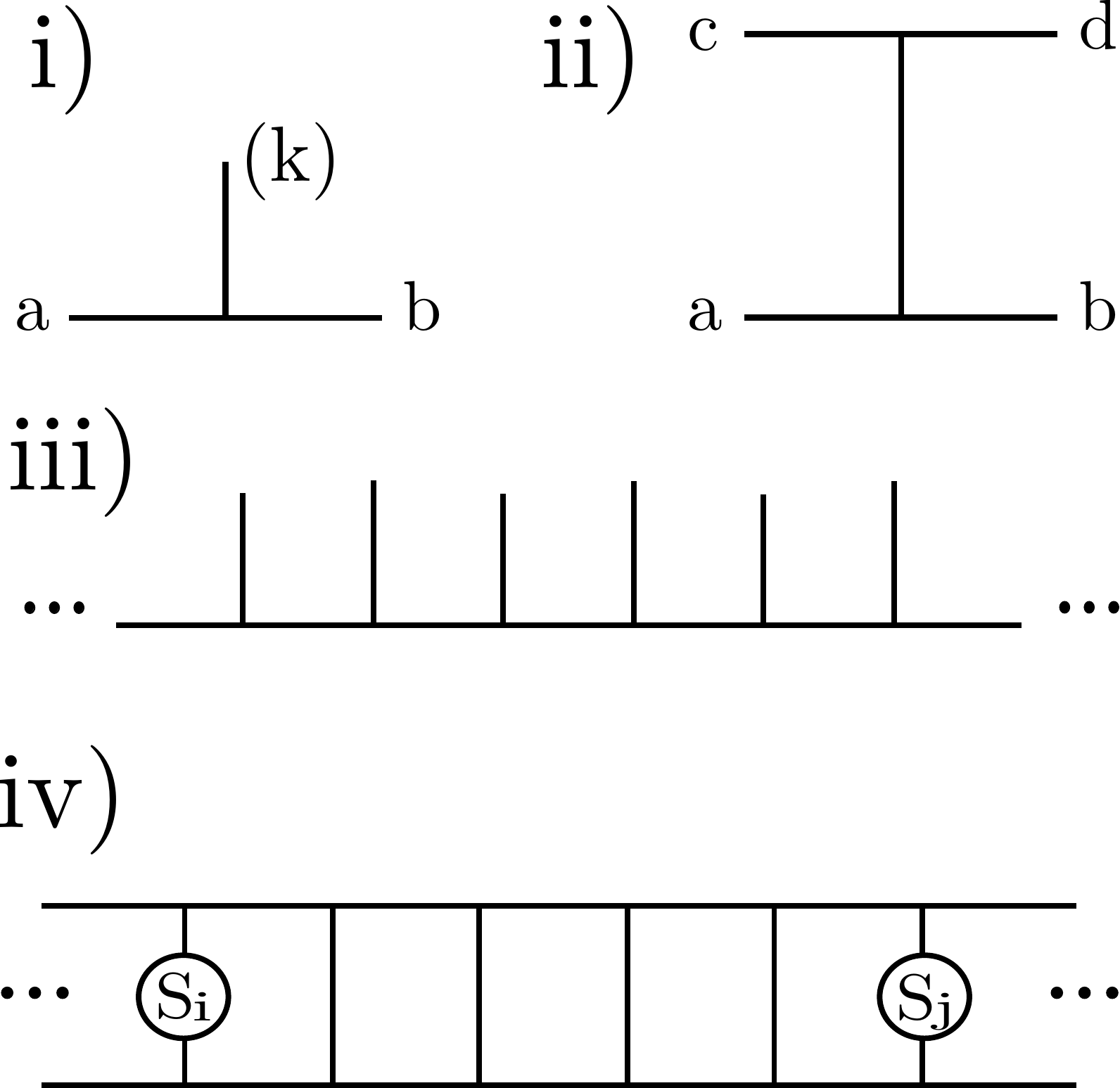}
 \caption{Pictorial representation of i) Matrix $M^{[k]}$ ii) One dimensional transfer matrix, iii)
 MPS Ground state, iv) Spin-spin correlation function. }
\label{fig:MPS}
\end{figure}

A convenient way to think of MPS is to represent them diagrammatically. We start by representing
the matrices $M$ as three legged objects (see Fig (\ref{fig:MPS}i)). The two horizontal lines are called
bonds and the state $|\Omega\rangle$ is obtained by contracting $N$ of this tensors along the horizontal
bonds (see Fig (\ref{fig:MPS}iii)). The computation of correlation functions becomes (assuming $j>i$)

\begin{equation}\label{correlation}
 \frac{\langle\Omega|S_i^aS_j^b|\Omega\rangle}{\langle\Omega|\Omega\rangle}=\frac{{\rm Tr}(E^{N-j+i-1}B_i(a)E^{j-(i+1)}B_j(b))}{{\rm Tr}E^N},
\end{equation}

\noindent where  $E=\sum_k\bar{M}^{[k]}\otimes M^{[k]}$ and $B_i(a)=\sum_{k_i,m_i}\bar{M}^{[k_i]}\otimes M^{[m_i]}\langle k_i|S_i^a|m_i\rangle $.
Pictorially $E$ is represented by Fig. (\ref{fig:MPS}ii). Due to the form of (\ref{correlation}) we call
$E$ the one dimensional transfer matrix. The correlation function is represented in Fig (\ref{fig:MPS}iv).

The MPS construction can describe {\it any} one dimensional state \cite{Verstraete2004,verstraete2006} for sufficiently large
matrices (whose size is called bond dimension).

In the VBS case, the one dimensional transfer matrix is $E_{(a,c);(b,d)}=\bar{M}^{[k]}_{ab}M^{[k]}_{cd}$ with

\begin{equation}  
E_{\rm VBS}=\begin{pmatrix}
  1 & 0 & 0 & 2\\
  0 & -1 & 0 & 0\\
  0 & 0 & -1 & 0\\
  2 & 0 & 0 & 1
 \end{pmatrix}.
\end{equation}

In the thermodynamic limit of infinite sites 
$\langle S_i^{a}S_j^{b}\rangle=\frac{4}{3}\left(-\frac{1}{3}\right)^{|i-j|}\delta_{ab}$. In \cite{Affleck1988},
the authors proved that (\ref{AKLT}) possess a gap in the spectrum in the thermodynamic limit. This is
in agreement with the exponential decay of correlation functions with distance.

\subsection{Entanglement and Renyi Entropies}

The previous constructions based on MPS are also useful to investigate the entanglement properties of ground states.
In the VBS case, this has been studied by many authors and in various contexts \cite{Fan2004,Xu2008,Katsura2010,Korepin2010}. Here we review 
some of these earlier results.

Entanglement between two subsystems of a pure state is characterized by its entanglement entropy (EE), that corresponds to
the von-Neumann entropy of one of the subsystems \cite{Amico2008}. It is defined by $ S_{\rm ent}=-{\rm Tr}(\rho_A\ln\rho_A),$
where $A$ is one of the subsystems. The partial density matrix is as usual
$\rho_A={\rm Tr}_{A_c}|\Omega\rangle\langle\Omega|/\langle\Omega|\Omega\rangle$ and $A_c$ is the complement of $A$. Note 
that for pure states $S_{\rm ent}(A)=S_{\rm ent}(A_c)$.
A generalization of EE that allows to obtain the eigenvalues of the partial density matrix is the Renyi
or $\alpha$-entropy, defined as 

\begin{equation}\label{Renyi}
 S_\alpha=\frac{\ln\rm Tr\rho_A^\alpha}{1-\alpha},
\end{equation}

\noindent in the limit $\alpha\rightarrow 1$, Renyi entropy becomes the entanglement entropy.

\begin{figure}[ht!]
\includegraphics[scale=.45]{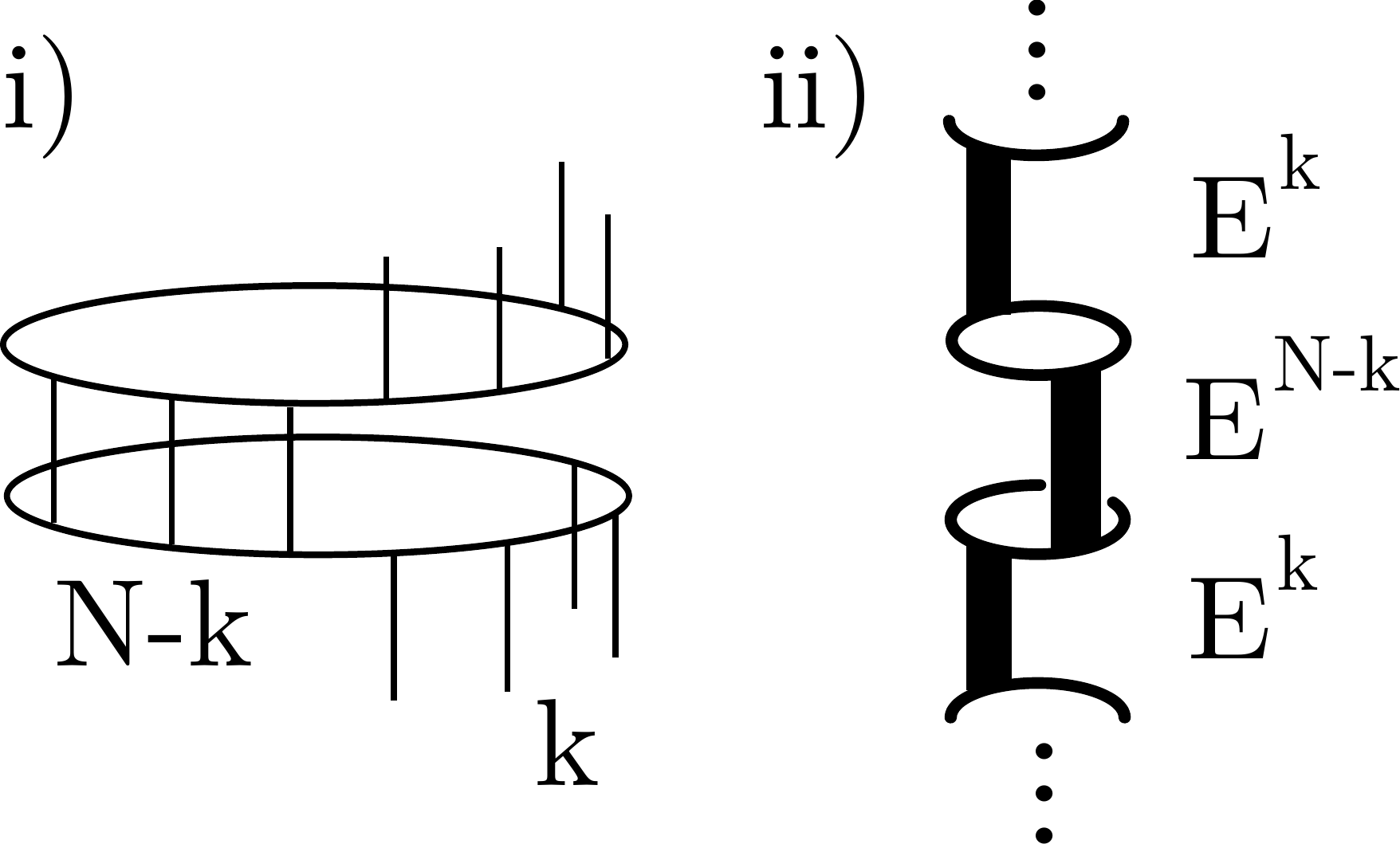}
 \caption{i) Partial density matrix for periodic boundary conditions. Here $k=4$ and $N=7$. ii) Contracting vertically
 $\rho_A$ $\alpha$-times we obtain $\rho_A^\alpha$. Thick black lines correspond to powers of transfer matrix $E$. }
\label{fig:reyni}
\end{figure}

Defining a partition of the spin chain into two blocks of length $k$ (subsystem $A$) and $N-k$ ($A_c$), the partial density
matrix is in general (using periodic boundary conditions)

\begin{equation}\label{trace}
 \rho_A=\bigotimes_{i,j=1}^k{\rm Tr}(E^{N-k}M^{[m_i]}\bar{M}^{[n_j]})|m_i\rangle\langle n_j|,
\end{equation}

\noindent where we have used the MPS representation for the ground state. Note that (\ref{trace}) contains $k$ matrices
$M$ and $\bar{M}$. Pictorially $\rho_A$ is shown in
Fig. (\ref{fig:reyni}i) . Contracting $\alpha$ of these objects in the vertical direction (i.e contracting the states
in the physical space) we obtain $\rho_A^\alpha$ (see also Fig (\ref{fig:reyni}ii)). 

\noindent The Renyi entropy is found to be \cite{Korepin2010, Santos2012b} (in the thermodynamic limit)

\begin{equation}\nonumber
 S_\alpha(k)=\frac{\ln[(1+3(-\frac{1}{3})^{k})^\alpha+3(1-(-\frac{1}{3})^{k})^\alpha]-2\alpha\ln2}{1-\alpha}.
\end{equation}

\noindent The EE in the double scaling limit ($k,N-k\rightarrow\infty$) is $S_{\rm ent}=2\ln 2$.
Here we find that EE saturates to a constant which is a consequence of the area law for EE of gapped systems in one dimension \cite{Hastings2007,Eisert2010}.

\section{Alternating bipartition and map to Eight Vertex Model}

\begin{figure}[ht!]
\includegraphics[scale=.43]{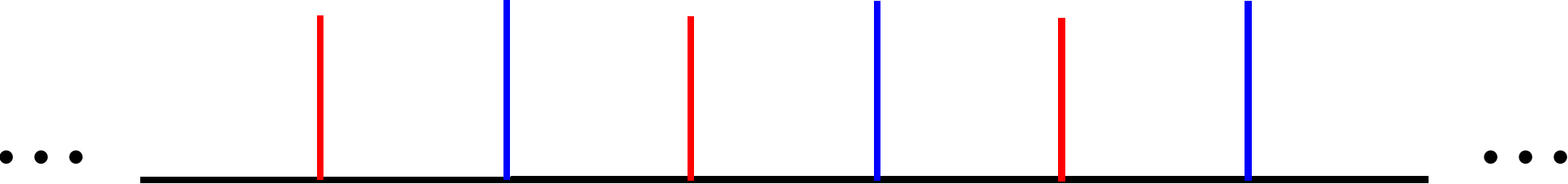}
 \caption{(Color online) Alternating bipartition. Sites in red belong to set $A$ while sites in blue
 belong to the complement set $A_c$.}
\label{fig:Alternating}
\end{figure}

In the previous section we reviewed some results about the entanglement of a bipartition defined by two
consecutive blocks of length $k$ and $N-k$.
In this section we proceed to investigate the entanglement of a different bipartition
defined as the partition between odd and even sites, that we call alternating bipartition (AB).
More specifically, let's number the sites in the chain from $0$ to $N-1$ for a chain of length $N$
and assume periodic boundary conditions. Sites at even positions belong to subset $A$ and sites
at odd positions belong to $A_c$ (See Fig. (\ref{fig:Alternating})). We assume the total number of sites $N$ to be even.

\begin{figure}[ht!]
\includegraphics[scale=.4]{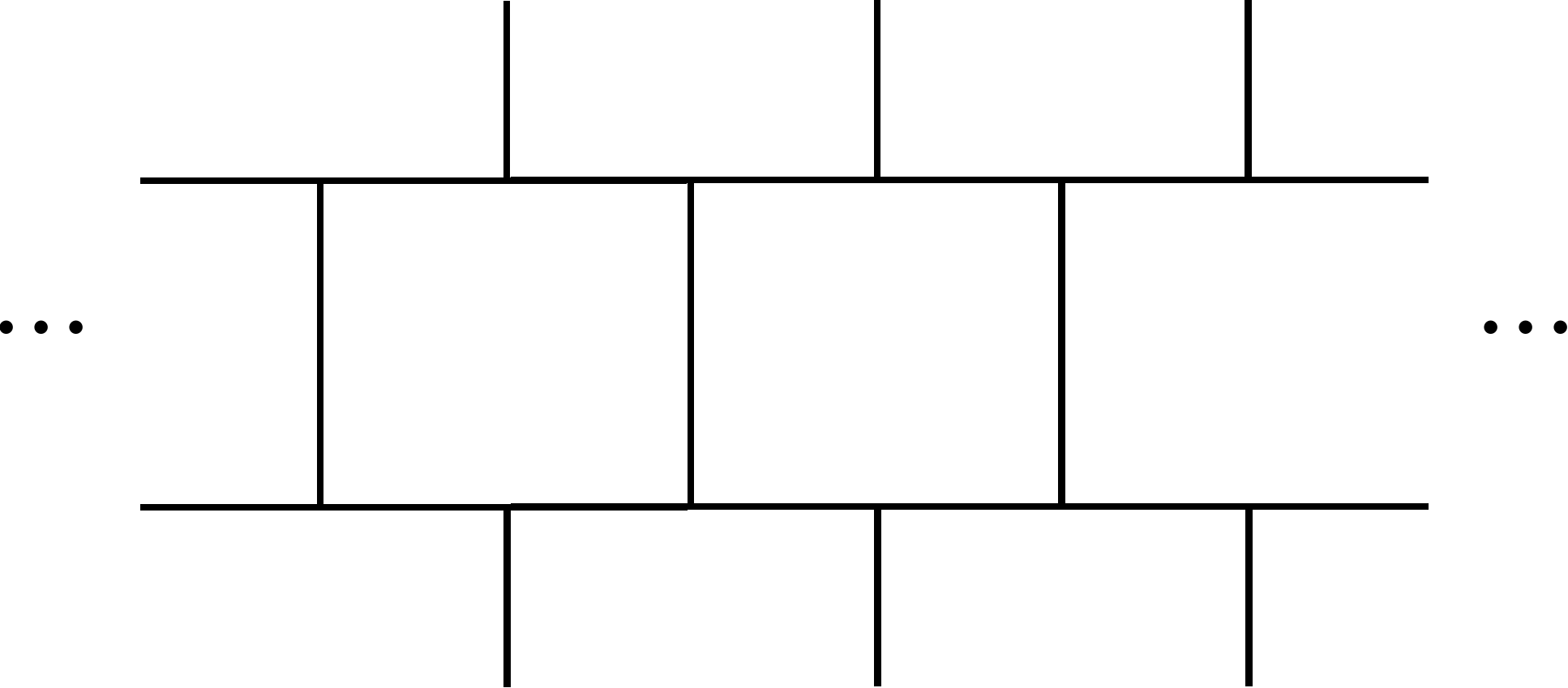}
 \caption{Partial density matrix for the alternating bipartition. This operator can be regarded as the transfer matrix of a two
 dimensional classical model.}
\label{fig:TM_AB}
\end{figure}

The partial density matrix $\rho_A$ is in this case given by Fig (\ref{fig:TM_AB}). To compute the Renyi entropy we have to stack $\alpha$ of these objects in the vertical direction. 
Taking the trace generates a two dimensional grid that can be
viewed as a two dimensional classical model.

In order to study the effective classical model defined by $\rho^\alpha$ can consider the more general transfer matrix ($u,v,w,t \in\mathbb{R}$)

\begin{equation}\label{TM}
  E=\bar{M}^{[k]}\otimes M^{[k]}
  =\begin{pmatrix}
  u & 0 & 0 & w\\
  0 & v & t & 0\\
  0 & t & v & 0\\
  w & 0 & 0 & u
 \end{pmatrix},
\end{equation}

\noindent which corresponds in general to the MPS of a spin $\frac{1}{2}$ ladder with groundstate $|\psi_0\rangle={\rm tr}(g_1g_2\dots g_{N-1}g_N),$
where 

\begin{equation}\label{MPS_time}
 g_i=\sum_k M^{[k]}|k\rangle=M^{[s]}|s\rangle+M^{[0]}|0\rangle+M^{[+]}|+\rangle+M^{[-]}|-\rangle.
\end{equation}

\noindent Here $|s\rangle$ corresponds to the singlet state along the rung, while $|-\rangle,|0\rangle,|+\rangle$ corresponds to one of the
triplet states with $S_z=m=-1,0,1$ respectively. The total $S_z$ measures the spin in the rung of the ladder. The explicit 
form of the $M^{[k]}$ matrices is 

\begin{eqnarray}\nonumber
M^{[s]} =\sqrt{\frac{u+v}{2}}\begin{pmatrix}
  1 & 0 \\
  0 & 1 
 \end{pmatrix},~
M^{[-]} =\sqrt{w}
 \begin{pmatrix}
  0 & \sin\theta \\
  -\cos\theta & 0 
 \end{pmatrix},\\\nonumber
 M^{[0]}=
 \sqrt{\frac{u-v}{2}}\begin{pmatrix}
  1 & 0 \\
  0 & -1 
 \end{pmatrix},~
M^{[+]} =\sqrt{w}
 \begin{pmatrix}
  0 &  \cos\theta \\
  -\sin\theta & 0
 \end{pmatrix}.
\end{eqnarray}

\noindent where $\sin 2\theta=-t/w$. For $u=-v=1,t=0$ and $w>0$ we recover the MPS (and consequently the transfer matrix) 
of the KSZ model \cite{klumper1992,klumper1993} ground state. Within the KSZ family, the point $w=2$ corresponds to the VBS 
state MPS.

In the case $u>|v|$ and $w>0$ the MPS (\ref{MPS_time}) is invariant under time reversal. The symmetry of this
MPS is enlarged to $SO(2)\sim U(1)$ (rotations in the xy plane) for $u=-v=1,t=0$ and $w>0$.

\begin{figure}[ht!]
\includegraphics[scale=.5]{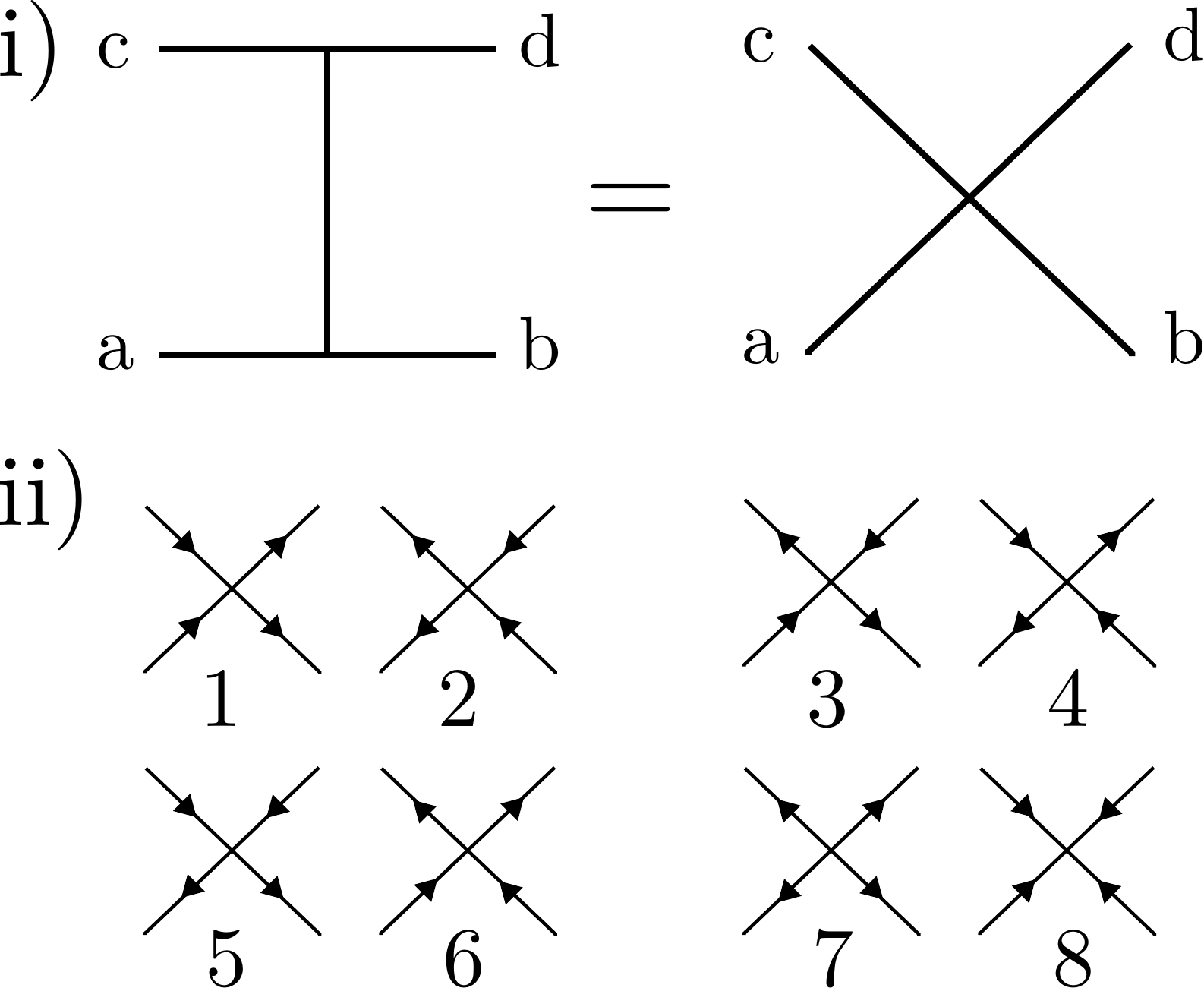}
 \caption{i) Transfer matrix as an arrow configuration. ii) Eight possible configurations with even
 number of arrows of the same type.In our notation, the arrows pointing North-East and South-East correspond to the label \textquoteleft $+$',
 North-West and South-West are assigned the label \textquoteleft $-$'}
\label{fig:Transfer_M}
\end{figure}

Mapping the transfer matrix (TM) four legged tensor to an arrow configuration (Fig. (\ref{fig:Transfer_M})), we can translate the computation of the Renyi 
($\alpha$) entropy to the partition function of the eight vertex model in a lattice with height $\alpha$ and width $N$, 
for a spin chain of length $N$. Specifically, defining the Boltzmann weights of the eight vertex model as $\omega_i$ for
the i-th arrow configuration, we establish (using $\pm\frac{1}{2}\rightarrow \pm$)

\begin{eqnarray}
 E_{(++);(++)}\equiv\omega_1,\quad E_{(+-);(+-)}\equiv\omega_3,\\
 E_{(+-);(-+)}\equiv\omega_5,\quad E_{(++);(--)}\equiv\omega_7.
\end{eqnarray}

\noindent Negating all the indices of $E$, and using that it has the form (\ref{TM}), we find that $\omega_1=\omega_2=u$, 
$\omega_3=\omega_4=v$, $\omega_5=\omega_6=t$ and $\omega_7=\omega_8=w$. These conditions correspond to the eight vertex model
in zero field \cite{Baxter8v1971}.

After taking the trace in (\ref{Renyi}), the Renyi entropy becomes equivalent (up to an 
overall constant) to the free energy of the eight vertex model in a lattice rotated $\pi/4$ radians (45 degrees) w.r. to the square lattice. This 
rotation is irrelevant for a classical system with open boundary conditions, but due to the periodic b.c. of the spin chain
and the trace in (\ref{Renyi}), we need to consider the classical partition function in a lattice wrapped on a torus.

The Renyi entropy is then

\begin{equation}\label{RE8VM}
 S_\alpha=\frac{\ln Z_{N,\alpha}(u,t,v,w)-N\alpha f_0(u,t,v,w)}{1-\alpha},
\end{equation}

\noindent where $N$ is the total number of sites and $Z_{N,\alpha}(u,t,v,w)$ is the eight vertex model partition function

\begin{equation}\label{part_func}
 Z_{N,\alpha}(u,t,v,w)=\sum_\mathcal{C}u^{n_1+n_2}v^{n_3+n_4}t^{n_5+n_6}w^{n_7+n_8},
\end{equation}

\noindent with $n_i$ the number of $i$ arrow configurations. The sum runs over all the allowed configurations of 
arrows $\mathcal{C}$ in the lattice. The last factor in (\ref{RE8VM}), $f_0$, which comes from the normalization of the 
partial density matrix, is given by

\begin{equation}
 f_0(u,t,v,w)=\ln\max{(|u\pm v|,|w\pm t|)},
\end{equation}

\noindent where the maximum has to be taken over all four possible choices of signs.

Note the lattice where the eight vertex model is defined in (\ref{part_func}) is rotated clockwise 45 degrees respect to
the usual square lattice orientation.

\subsection{Reduction to a Rotated Six Vertex Model}

To make further progress we use the symmetry of the partition function (\ref{part_func}). For $N$ even, 
(which is the case we are considering), the lattice can be divided into two sublattices $S_1$ and $S_2$ such that
every vertex in $S_1$ has neighbors just in $S_2$, and viceversa. Reversing all arrows on the edges that start on a site
in $S_1$ and go south west or south east, we get a new zero field eight vertex model, but with $u,t,v,w$ replaced by
$v,w,u,t$, so

\begin{equation}
 Z_{N,\alpha}(u,t,v,w)=Z_{N,\alpha}(v,w,u,t)
\end{equation}

In the VBS case, and in its KSZ generalization, we have $t=0$. Focusing in this case, 
$Z_{N,\alpha}(v,w,u,0)$ corresponds to the partition function of the six vertex model solved in \cite{Lieb1967}.
This connection between the KSZ MPS and the six vertex model can be further understood comparing the symmetries
of these two systems. The six vertex effective Hamiltonian (defined as the logarithmic derivative
of the transfer matrix operator) corresponds to the XXZ model which has $U(1)$ symmetry (rotations in the xy plane), 
as it is also the KSZ transfer matrix (\ref{TM}).

\subsection{Alternating Entanglement of VBS as a CFT}

As discussed previously, for the MPS states studied here, the Renyi entropy of the entanglement bipartition can be mapped to a classical vertex model.

\begin{figure}[ht!]
\includegraphics[scale=.55]{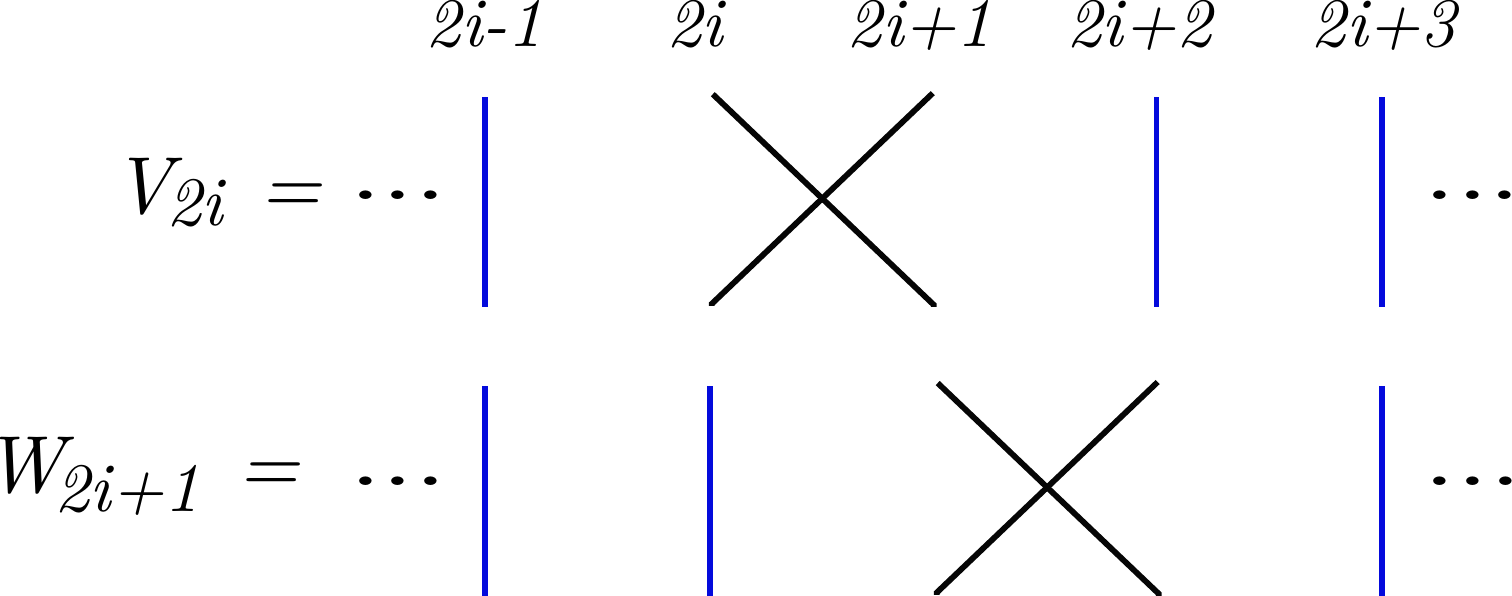}
 \caption{Operators $V_{2i}$ and $W_{2i+1}$. $V_{2i}$ ($W_{2i+1}$) acts nontrivially just in adjacent spaces $2i$ and $2i+1$ 
 ($2i+1$ and $2i+2$). In these spaces $V_{2i}$ ($W_{2i+1}$) acts as the transfer matrix $E$. Each blue line represents 
 the identity in the corresponding space.}
\label{fig:VW}
\end{figure}

In this section we will show that the VBS case is special as it maps to a critical point in the vertex model parameter space.
A simple way to see this connection is the following. We introduce the operators (see also Fig. \ref{fig:VW})

\begin{eqnarray}\label{matrices}
 V_{2i}&=&e_0 \dots \otimes e_{2i-1}\otimes E_{2i}\otimes e_{2i+2}\dots\otimes e_{N-1},\\
 W_{2i+1}&=&e_0 \dots \otimes e_{2i}\otimes E_{2i+1}\otimes e_{2i+3}\dots\otimes e_{N-1},
\end{eqnarray}

\noindent with $E_{i}$ the one dimensional transfer matrix acting at space $i$ and $i+1$. Here $e_i$ is a $2\times 2$ identity 
matrix at site space $i$. 

The transfer matrix of the two dimensional system (see Fig \ref{fig:TM_AB}) is then given by $\mathcal{VW}$, where

\begin{equation}
 \mathcal{V}=\prod_{i=0}^{(N-2)/2}V_{2i}\quad\mbox{and}\quad\mathcal{W}=\prod_{i=0}^{(N-2)/2}W_{2i+1}.
\end{equation}

\begin{figure}[ht!]\cite{Baxter_Book}
\includegraphics[scale=.43]{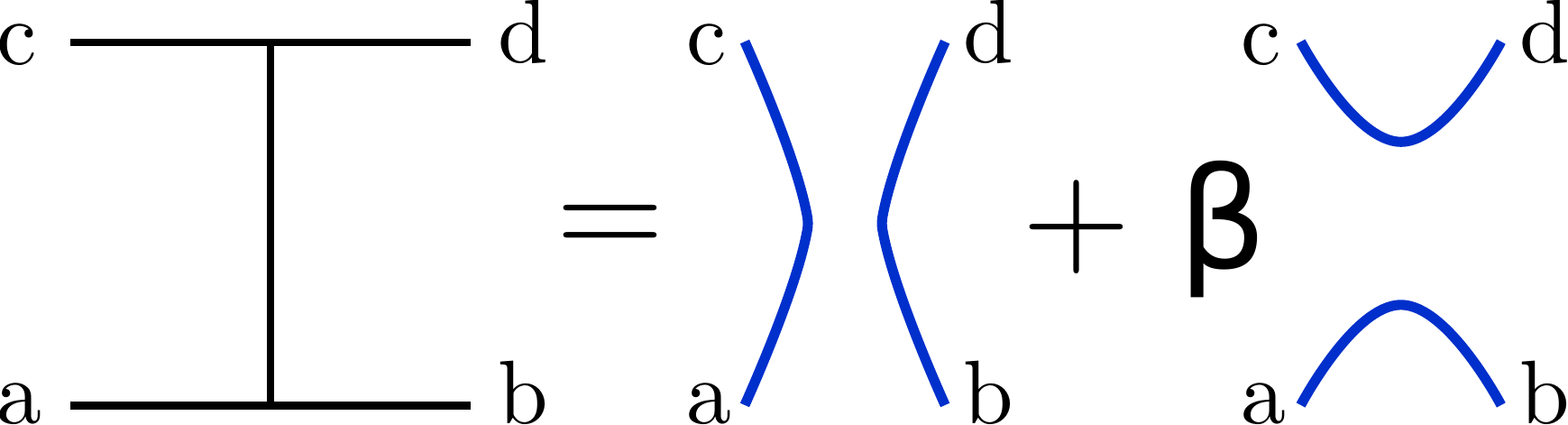}
 \caption{i) Transfer matrix as linear combination of Temperley-Lieb generators.
 ii) Graphical definition of operator $U_{i}$. It acts nontrivially just in the spaces $i$ and $i+1$.  
 In the top and bottom figures, each blue line represents a Kronecker $\delta$ function}
\label{fig:Transfer_Lieb}
\end{figure}

Plugging parametrization of the transfer matrix (see Fig. \ref{fig:Transfer_Lieb}i)

\begin{equation}\label{TM_Temperley}
E_{(ac);(bd)}=\delta_{ac}\delta_{bd}+\beta\delta_{cd}\delta_{ab},
\end{equation}

\noindent on the definitions of $\mathcal{V}$ and $\mathcal{W}$  (\ref{matrices}),
we have ($\beta=-\frac{1}{2}$ for the VBS case)

\begin{equation}
\mathcal{V}=\prod_{i=0}^{\frac{N}{2}-1}(I+\beta U_{2i}),\quad  \mathcal{W}=\prod_{i=0}^{\frac{N}{2}-1}(I+\beta U_{2i+1}).
\end{equation}

The operators $U_0,\dots U_{N-1}$ are defined diagrammatically in Fig. (\ref{fig:Transfer_Lieb}ii). 
They satisfy the Temperley-Lieb (TL) algebra \cite{TLalgebra}

\begin{eqnarray}\label{TLalg}
 U_{i}^2&=&\sqrt{q}U_{i},\\
 U_{i}U_{i+1}U_{i}&=&U_i,\\
 U_iU_j&=&U_jU_i, \quad |i-j|\geq 2.
\end{eqnarray}

\noindent with $q=4$. This is a particular representation of the TL algebra. As noted in \cite{TLalgebra},
an equally good representation of the algebra can be achieved by the $2^N\times 2^N$ matrices

\begin{eqnarray}
 (U_{2i-1}^{R})_{\sigma,\sigma'}&=&q^{-\frac{1}{2}}\prod_{j=1\neq i}^{N}\delta_{\sigma_j,\sigma'_j},\\
 (U_{2i}^{R})_{\sigma,\sigma'}&=&q^{\frac{1}{2}}\delta_{\sigma_i,\sigma_{i+1}}\prod_{j=1}^{N}\delta_{\sigma_j,\sigma'_j}.
\end{eqnarray}

\noindent Writing down the matrices $\mathcal{V}$ and $\mathcal{W}$ as before, but with this representation of the matrices $U$, 
they can be expressed as \cite{TLalgebra, Baxter_Book}

\begin{eqnarray}
 \mathcal{V}^R_{\sigma,\sigma'}&=&\exp\left(K_1\sum_{j=1}^{N-1}\delta_{\sigma_j,\sigma_{j+1}}\right)\prod_{j=1}^N{\delta_{\sigma_j,\sigma_j'}},\\
 \mathcal{W}^R&=&\exp\left(K_2\sum_{j=1}^{N}\delta_{\sigma_j,\sigma_{j}'}\right).
\end{eqnarray}

\noindent with $K_1=\ln(1+\beta\sqrt{q})$ and $K_2=\ln(1+\frac{1}{\beta\sqrt{q}})$. The matrices $\mathcal{V}^R,\mathcal{W}^R$ 
build up the partition function of the Potts models at the criticality. The critical manifold is defined when $(e^{K_1}-1)(e^{K_2}-1)=q$ 
(note that this condition is independent of the parameter $\beta$ in (\ref{TM_Temperley})). For
the VBS case, $q=4$ and $\beta=-1/2$, so the corresponding Potts model is ferromagnetic, critical and at zero temperature.
Given that the partition function  does not depend on the particular
representation of the TL generators \cite{TLalgebra}, we find that the Renyi entropy is 
effectively described by a conformal field theory of central charge $c=1$ \cite{Baxter_Book, Belavin1984, Blote1986, Fujimoto1996}
which describes the Potts model at criticality for $q=4$ (for $N\rightarrow\infty$ and $\alpha\gg 1$).

\subsection{Generalization to $SU(n)$}

A natural generalization of the previous discussion is to consider $SU(n)$ invariant VBS states,
where $n=2$ represents the original AKLT ground state. Here we review the construction of such states
following \cite{Katsura2008, Korepin2010}.
At each lattice site of a chain, we consider particles in the adjoint representation of $SU(n)$ (in the $SU(2)$ case
this corresponds to spin 1 particles). By analogy with the original AKLT case, we construct such representations
by considering the tensor product of the fundamental and its conjugate representation (antifundamental) of $SU(n)$ in each 
site, $\Box\otimes\overline{\Box}=$ singlet $\oplus$ adjoint, and projecting this product on the adjoint irrep subspace,
(similar to constructing spin 1 irrep out of two spin $\frac{1}{2}$ in the original case). To complete the construction
we form singlets between nearest neighbor pairs of fundamental and conjugate representations.
Choosing a basis for the states in the fundamental representation $|a\rangle$ ($a=1,\dots,n$) and $|\bar{a}\rangle$ for states
in the conjugate, the projector onto the adjoint representation can be written as

\begin{equation}\label{TM_SUN}
\mathcal{P}_{\rm adj}=\sum_{a,\bar{a}}|a,\bar{a}\rangle\langle a,\bar{a}|-\frac{1}{n}\sum_{a,b}|a,\bar{a}\rangle\langle b,\bar{b}|,
\end{equation}

\noindent where we have used that the representation of $\Box\otimes\overline{\Box}$ splits into the singlet and
adjoint irreps. The projector into the singlet state is just $|0\rangle\langle 0|$ with 
$|0\rangle= \frac{1}{\sqrt{n}}\sum_{a}|a,\bar{a}\rangle$ being the singlet.

The transfer matrix for the $SU(n)$ symmetric state described above is given by 
$E=\mathcal{P}_{\rm adj}^\dagger\mathcal{P}_{\rm adj}$ \cite{Orus2011}. Using (\ref{TM_SUN}), it is easy to see
that $E=\mathcal{P}_{\rm adj}$. 
The matrix elements of $E$ read

\begin{equation}
 \langle a, \bar{b}|E|c,\bar{d}\rangle=\delta_{ac}\delta^{bd}-\frac{1}{n}\delta_{a}^b\delta_c^d
\end{equation}

\noindent where the upper indices corresponds to the conjugate representation \footnote{In \ref{TM_Temperley}, we do not 
distinguish between upper an lower indices, as the fundamental representation and its conjugate are equivalent for $SU(2)$}.

\begin{figure}[ht!]
\includegraphics[scale=.4]{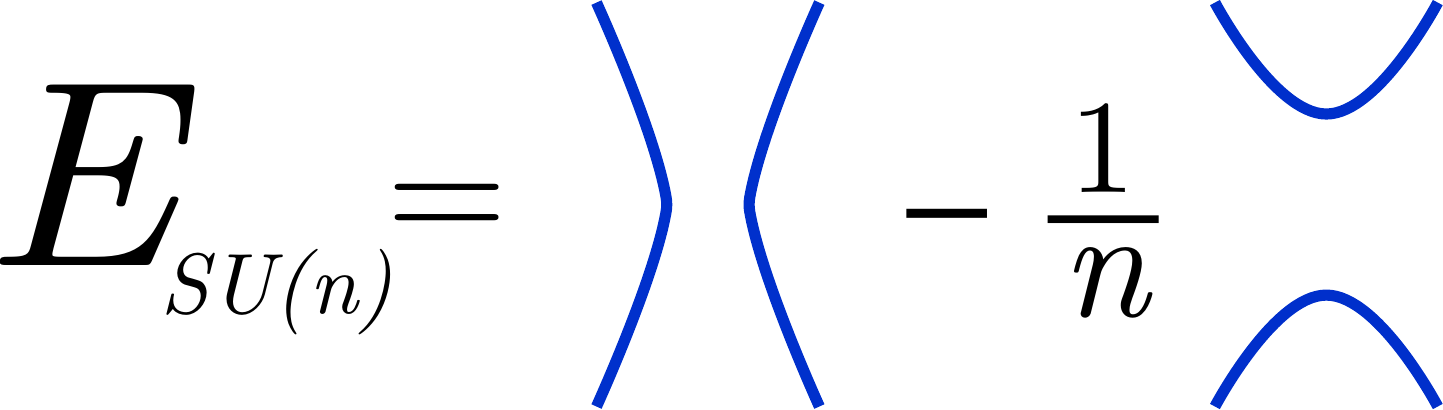}
 \caption{(color online). Graphical representation of transfer matrix $E$ for $SU(n)$ VBS. Each blue line represents the action 
 of the $n\times n$ identity in the corresponding space}
\label{fig:Transfer_SUN}
\end{figure}

As a mapping between vector spaces, the matrix $E$ can be represented as Fig. (\ref{fig:Transfer_SUN}). From the discussion 
in the previous subsection, it is clear that this transfer matrix generates operators $U_i$ satisfying
the Temperley-Lieb algebra (\ref{TLalg}) with $q=n^2$.

Following the arguments of the previous subsection, the Reyni entropy of the AB of $SU(n)$ VBS maps
to a critical $n^2$-state Potts model. It is interesting to note that the $q-$state Potts model for $q>4$ at 
criticality undergoes a {\it first} order phase transition. This means that there is \textbf{no} conformal field theory 
description for the alternating entanglement for this generalized $SU(n)$ state with $n>2$. 

Some insight into this transition can be obtained studying the effective Hamiltonian obtained from the logarithmic 
derivative of the 2D transfer matrix. This Hamiltonian inherits the symmetries of the transfer matrix $E$. In the 
cases considered here, the effective Hamiltonian should possess $SU(n)$ symmetry. 
For $n=2$, the effective Hamiltonian corresponds to the Heisenberg $XXX$ model. 
This model is exactly solvable by Bethe ansatz and possess a unique ground state. In the case $n>2$,
$SU(n)$ antiferromagnetic chains possess a dimerized ground state with a finite gap \cite{Affleck1990}.

The first order transition in the AB entanglement is a signature of this dimerization transition in the 
effective Hamiltonian, which is essentially the entanglement Hamiltonian. The details of this connection will be published 
elsewhere \cite{Raul2}.

\subsection{Topological point}

\begin{center}
\begin{figure}[ht!]
\includegraphics[scale=.45]{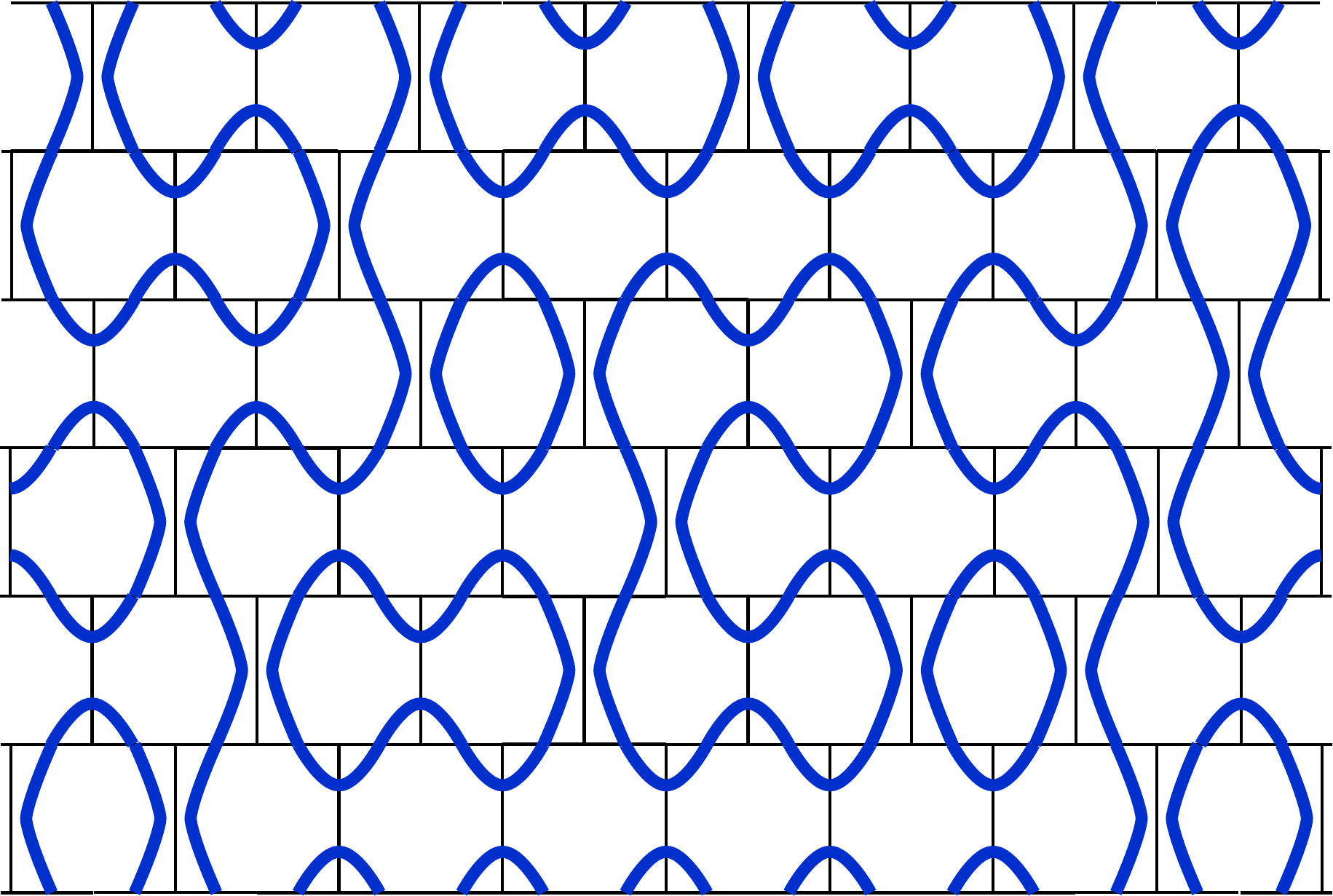}
 \caption{Loop covering for the Renyi entropy of an alternating bipartition. Here, the number of sites is 8 and $\alpha=6$.
 Periodic boundary conditions are assumed}
\label{fig:Loops}
\end{figure}
\end{center}

The one dimensional transfer matrix (\ref{TM_Temperley}) at the point $\beta=1$ describes a
systems whose partition function is not only critical but also topological. The associated
classical model becomes a sum over loop configurations in the torus (see Fig (\ref{fig:Loops})). The partition function 
in this case can be written as

\begin{equation}
 Z=\sum_\mathcal{C}q^{N(\mathcal{C})/2}.
\end{equation}

\noindent Here the sum is performed over loop configurations, where $N(\mathcal{C})$ is the total number of loops in the
configuration $\mathcal{C}$. This point does not correspond to the ground state of the KSZ model for any value of the 
parameters.

The Hamiltonian whose ground state corresponds to this MPS can be expressed as a local operator acting on
spins $\frac{1}{2}$ ladders \cite{Raul2}. It is important to note that the topological property of the partition 
function is present for any $\alpha$, i.e. any height of the classical model.

Interestingly, in the $SU(n)$ generalization, the topological point corresponds a group of negative dimension $n=-1$. 
Groups with negative dimension can be properly defined in the context of tensor categories \cite{Penrose}.

\section{Conclusions}

In this letter we have analyzed the Reyni entropy of entanglement for an alternating bipartition (AB) of MPS states in one 
dimension. This partition assigns nearest neighbor sites to different complementary sets $A$ and $A_c$.
We show the equivalence of the AB Renyi entropy, with the partition function of an eight vertex model for a 
generic class of MPS states invariant under time reversal. At particular values of the parameters, 
the MPS possess a $U(1)$ symmetry and describes the ground state of the KSZ model. In this case the Renyi entropy 
can be further mapped to a six vertex model. At the fully $SU(2)$ symmetric point, corresponding to the VBS state, we showed
that the classical model describing the Renyi entropy is critical, with central charge $c=1$. 

In the $SU(n)$ generalization of the VBS construction
we find that the classical model corresponds to a $n^2$-state Potts model at criticality. For $q>4$ the $q$-state Potts
model transition is of first order, signaling in this context a dimerization transition in the entanglement
Hamiltonian for the $SU(n)$ VBS ground state.

We also discuss a special point where the corresponding classical model is not only conformal but also
topological. It maps to a loop configuration, which is a quantity that depends on the topology of the lattice. The 
topological partition function that describes the Reyni ($\alpha$) entropy remains topological even in the 
limit $\alpha\rightarrow 1$, where the Reyni entropy becomes the von-Neumann entropy of entanglement.

Although we have focused in one dimensional systems, the mapping between AB Renyi entropy of a $D$ dimensional quantum system
and a classical partition function in $D+1$ dimensions is general.

\smallskip

\textbf{Acknowledgments} R. S. is glad to thank the organizers of {\it Condensed Matter in the City} workshop, 
where this work was motivated. The author acknowledges P. Coleman, V. Korepin, and T. Wei for useful discussions. 
The author would also like to thank the anonymous referee for his/her valuable comments. 
R.S. is supported by grant GIF 1167-165.14/2011.

\smallskip

{\it Note added:} Upon completion of this work we noticed the results obtained recently in \cite{Rao2014,Hsieh2014b}
which overlaps with some of the results obtained here. Our result confirms the numerical prediction for the central
charge of the associated classical model for the VBS state.


%

\end{document}